\documentclass[manuscript,screen]{acmart}
\AtBeginDocument{%
  }

\setcopyright{acmlicensed}
\copyrightyear{2018}
\acmYear{2018}
\acmDOI{XXXXXXX.XXXXXXX}
\acmConference[Nanyang Blockchain Conference 2025]{Yotta: A Large-Scale Trustless Data Trading Scheme for Blockchain System}{Aug 16-17, 2025}{Singapore}
\acmISBN{978-1-4503-XXXX-X/2018/06}

\usepackage[textsize=tiny,textwidth=0.6in]{todonotes}
\newcommand{\allnotes}[1]{}
\renewcommand{\allnotes}[1]{#1}

\begin{document}

\title{Yotta: A Large-Scale Trustless Data Trading Scheme for Blockchain System}

\author{Xiang Liu}
\authornote{Both authors contributed equally to this research. The order of the two authors was chosen randomly.}
\email{liuxiang@comp.nus.edu.sg}
\orcid{0009-0006-8550-3767}
\affiliation{
  \institution{National University of Singapore}
  \city{Singapore}
  \country{Singapore}
}

\author{Zhanpeng Guo}
\authornotemark[1]
\email{zhanp.guo@gmail.com}
\orcid{0009-0000-8722-774X}
\affiliation{%
  \institution{Xidian University}
  \city{Xi'an}
  \country{China}
}
\author{Liangxi Liu}
\email{liu.liangx@northeastern.edu}
\orcid{0009-0000-4074-5144}
\affiliation{
  \institution{Northeastern University}
  \city{Boston}
  \country{USA}
}

\author{Mengyao Zheng}
\email{mengyaozheng@alumni.harvard.edu}
\orcid{0009-0006-8514-9059}
\affiliation{
  \institution{Harvard University}
  \city{Boston}
  \country{United States}
}
\author{Yiming Qiu}
\email{yimingq@umich.edu}
\orcid{0009-0003-9328-3205}
\affiliation{
  \institution{University of Michigan}
  \city{Ann Arbor}, 
  \country{United States}
}


\author{Linshan Jiang}
\authornote{Corresponding Author.}
\email{linshan@nus.edu.sg}
\orcid{0000-0001-8501-9488}
\affiliation{
  \institution{National University of Singapore}
  \city{Singapore}
  \country{Singapore}
}

\renewcommand{\shortauthors}{Trovato et al.}

\begin{abstract}
Data trading is one of the key focuses of Web 3.0. However, all the current methods that rely on blockchain-based smart contracts for data exchange cannot support large-scale data trading while ensuring data security, which falls short of fulfilling the spirit of Web 3.0. 
Even worse, there is currently a lack of discussion on the essential properties that large-scale data trading should satisfy. 
In this work, we are the first to formalize the property requirements for enabling data trading in Web 3.0. Based on these requirements, we are the first to propose \textbf{Yotta} — a complete batch data trading scheme for blockchain — which features a data trading design that leverages our innovative cryptographic workflow with IPFS and zk-SNARK. Our simulation results demonstrate that Yotta outperforms baseline approaches up to 130 times and exhibits excellent scalability to satisfy all the properties.
\end{abstract}

\ccsdesc[500]{Security and privacy~Distributed systems security}
\ccsdesc[500]{Security and privacy~Information accountability and usage control}
\ccsdesc[500]{Networks~Peer-to-peer protocols}

\keywords{Blockchain, Data Trading, Web 3.0}

\received{20 February 2007}
\received[revised]{12 March 2009}
\received[accepted]{5 June 2009}

\maketitle

\section{Introduction}
We are transitioning from the Web 2.0 era to the Web 3.0 era.
In Web 3.0, AI has become a rapidly developing focus. For AI, data and computing power have, in fact, become the most valuable resources, with data serving as the foundation.
As a result, data trading has emerged as one of the key focuses of Web 3.0~\cite{jadbabaie2014ieee}.

However, the current methods for data trading on Web 3.0 data markets remain limited. For instance, trading AI datasets or medical data at a large scale often requires going through official authorities, which is complex and cumbersome. Although such processes can mitigate the risks of deception by untrusted parties, they contradict the core principles of Web 3.0 — decentralization~\cite{mozilla2024internet}, User ownership/sovereignty over data~\cite{krause2024web3}, trustless systems~\cite{opencamp2025web3talk}, and permissionless innovation~\cite{entin2023web3} — and ultimately hinder the development of Web 3.0.

On the contrary, in Web 3.0, everyone can freely contribute and trade data from themselves, and the volume of data has already reached the zettabyte scale~\cite{idc2018datasphere}. Although using Ethereum smart contract~\cite{wood2014ethereum} on Web 3.0 can ensure the security of data, it still fails to meet practical demands. First, it cannot support large-scale data purchasing, such as many-to-many transactions in real-world scenarios. Second, the traded data is trustless, making it difficult to assess the credibility of the entities or organizations involved, which can potentially result in the exchange of garbage data and expose participants to significant financial loss risks.


Even worse, beyond just security and large-scale data trading, there is currently no formalized summary of the properties that an ideal scheme for large-scale data trading in Web 3.0 should satisfy. A natural question then to ask is "What should an ideal scheme look like?" To address this gap, we are the first to propose the ``\textbf{SQUATS}" principle that an ideal data trading scheme should fulfill:

\begin{itemize}
    \item \textbf{Scalability}: The scheme should not only support atomic transactions but also enable large-scale batch data trading as well as many-to-many transactions.

 \item \textbf{Quality}: The scheme should allow verification of data quality, as high-quality data is fundamental for data mining and machine learning. It should ensure that the seller's data meets the buyer's specified requirements.

 \item \textbf{Usability}: The scheme should be easy to use, user-friendly, and extensible for broader adoption.

 \item \textbf{Autonomy}: The scheme should enable automated trading, integrating seamlessly with existing popular smart contracts without excessive manual intervention or delays. 
 

 \item \textbf{Transparency}: The platform should be transparent, eliminating the need for a trusted third party. Buyers and sellers should not have to rely on intermediaries to escrow data or payments.

 \item \textbf{Security}: The scheme should guarantee data security, preventing malicious third parties from tampering with or decrypting the data. It should also ensure that sellers provide the correct decryption keys, and that only authorized buyers can access the purchased data.
\end{itemize}

Building on the above principle, we are the first to integrate the spirits of Web 3.0 and provide a complete batch data trading scheme based on blockchain, named \textbf{Yotta}. 
Yotta leverages Inter-Planetary File System (IPFS)~\cite{benet2014ipfs}, zero-knowledge succinct non-interactive argument of knowledge (zk-SNARK)~\cite{gabizon2019plonk,chen2023hyperplonk}, and novel cryptographic algorithms and workflow designs to realize an innovative encrypted data trading scheme for blockchain. The key of Yotta lies in secure and reliable large-scale batch data trading and satisfies the aforementioned properties. Our method demonstrates excellent scalability, paving the way for handling future yottabyte scale data volumes. Furthermore, Yotta is designed to be easily horizontally extensible, allowing seamless integration with future methods for enhanced performance. Our contributions are summarized as follows:

\begin{itemize}
\item We are the first to systematically propose the essential properties that a large-scale data trading scheme in Web 3.0 should satisfy, and we summarize them as the \textbf{SQUATS} principles.

\item Based on these principles, we propose \textbf{Yotta}, a blockchain-based data trading scheme that satisfies all the identified properties. Yotta is also designed to be extensible and can be easily integrated with existing systems.

\item We build our Yotta prototype and conduct simulation experiments, demonstrating that our approach achieves up to 130× acceleration compared to baseline methods.
\end{itemize}

\section{Related Works}
\subsection{Key Elements in Web 3.0}

\subsubsection{Ethereum}
The Ethereum protocol was the first to implement practical smart contracts, extending the Bitcoin system~\cite{nakamoto2008bitcoin} and becoming one of the cornerstones of blockchain technology. Due to Ethereum's scalability and versatility, its role has evolved beyond merely supporting digital currencies, such as Decentralized Finance (DeFi) and Non-Fungible Tokens (NFTs), to enabling decentralized applications (DApps). Ethereum allows a large number of entities to participate in the execution of smart contracts, with each maintaining a copy of the blockchain ledger. This decentralized nature ensures transparency and immutability, enabling the realization of atomic transactions. Our Yotta prototype is also built upon improvements to Ethereum smart contracts.

\subsubsection{zk-SNARK}
Zero-knowledge proofs (ZKPs), especially zero-knowledge succinct non-interactive arguments of knowledge (zk-SNARK), have been widely adopted in the Web 3.0 ecosystem—not only in cryptocurrencies~\cite{bowe2020zexe}, but also in blockchain rollups~\cite{liu2024pianist}, AI regulations~\cite{liu2021zkcnn}, and anonymous credentials~\cite{rosenberg2023zk}. ZKPs offer strong privacy guarantees, support scalability, and eliminate the need for trusted third parties, aligning with the principle of minimal disclosure. zk-SNARKs, in particular, are attractive due to their succinct constant-sized proofs and verification complexity that remains independent of the data size, making them highly applicable in real-world systems. In fact, any problem with complexity below or equivalent to NP-complete can be efficiently reduced to a zk-SNARK-friendly format. As a result, Yotta integrates zk-SNARK to enable verifiable data quality and reduce the overall verification complexity.

\subsubsection{IPFS}
Libp2p allows users to discover peers and establish seamless communication through a Distributed Hash Table (DHT), while also supporting decentralized content distribution. As a foundational technology within the Web 3.0 ecosystem, IPFS integrates libp2p to build its peer-to-peer network. IPFS is a distributed, decentralized file sharing system, where file contents are located and verified using cryptographic hashes to ensure data integrity. Consequently, a large number of users in Web 3.0 and cloud services leverage IPFS to share their content. In our Yotta prototype, we also adapt IPFS as the underlying system for data storage.

\subsection{Data Trading Schemes}


Early data trading schemes primarily focused on privacy. Zyskind et al.~\cite{zyskind2015decentralizing} first propose a decentralized approach for personal privacy protection. Yue et al.~\cite{yue2017big} focus on security to explore blockchain-based models for big data sharing. Zheng et al.~\cite{zheng2017overview} further provide a comprehensive review of blockchain applications in large-scale data sharing. Dong et al.~\cite{dong2018efficient} leverage secure multi-party computation and differential privacy to enhance data protection, enabling potential blockchain applications. However, these approaches lack discussions with practically supporting large-scale data trading.

Although Jung et al.~\cite{jung2017accounttrade,jung2018accounttrade} support large-scale data markets, their schemes assume that agents are trusted, which can lead to information leakage. Dai et al.~\cite{dai2019sdte} propose a traceable data trading protocol, but it does not guarantee data quality. Taleb et al.~\cite{taleb2018big} provide a quality evaluation framework but without sufficient implementation details. Zheng et al.~\cite{zheng2020blockchain} are the first to explore large-scale data trading on the blockchain; however, their prototype remains inefficient and cannot effectively support one-to-many transactions with a trusted key center.
Some researchers have further explored the use of techniques such as Divisible Computation Diffie–Hellman (DCDH) as their decentralized method to exchange data in a manner similar to key exchange without a trusted third party to fulfill the requirements from buyer and seller in Web 3.0. Nevertheless, these approaches still face similar limitations: they struggle to achieve scalability when integrated with blockchain and smart contracts for large-scale data trading.

We are the first to ensure privacy while supporting large-scale data exchange, and simultaneously satisfying all properties defined in our SQUATS principles. In the next section, we will describe our scheme in detail, starting with an introduction to how Yotta achieves one-to-many transactions as a foundation for realizing many-to-many transactions.

\section{Our Methods}



\subsection{System Model}
Our Yotta system model mainly consists of three entities: Buyer, Seller and Blockchain system.
\begin{itemize}
    \item Buyer: Intends to purchase data from one or more sellers.
    \item Sellers: Multiple parties (sellers) who own and wish to sell their respective data, which is stored on IPFS.
    \item Blockchain: Responsible for running smart contracts and verifying and executing the payment transactions.
\end{itemize}

\subsection{Preliminaries}
This subsection introduces the mathematical notations used in this paper. The following examples use IPFS for illustration purposes; however, any data platform could be used in practice. Our current prototype is implemented on IPFS.

\begin{itemize}
\item For the data trading, we have dataset $(Dataset, D)$. The dataset owned by the $n$ sellers consists of multiple secret data items $[S_1, S_2,\cdots S_i \cdots S_n]$, where each $S_i$ represents the data belonging to seller $i$ and all the data is stored on IPFS.

\item IPFS Content Hash $(IPFS Hash,  H{ipfs})$: Used to uniquely identify data stored on IPFS.

\item zk-SNARK Proof $(Proof, \pi)$: Each seller must generate a proof demonstrating that they indeed possess each secret item in the dataset, and that these items match the data referenced by their corresponding IPFS addresses.

\item Encryption and Decryption Function (either symmetric or asymmetric) $(Encrypt, Decrypt)$: Used to encrypt and decrypt each address for trading data.

\item The seller discloses a subset of data samples along with their labels. The buyer examines the samples to verify that they meet the intended requirements, and publicly specifies an evaluation function $F()$.
\end{itemize}

\begin{figure}[!t]
\centering
\includegraphics[width=\linewidth]{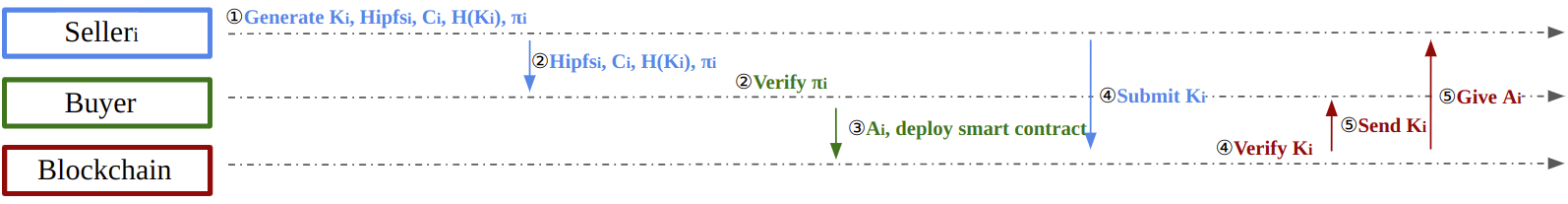}
\caption{Overview of our Yotta scheme, which consists of five steps.}
\label{fig:scheme}
\end{figure}

\subsection{Yotta Scheme}

Our Yotta Scheme is primarily composed of five steps, as shown in Figure~\ref{fig:scheme}. Then, we will discuss the five steps in detail in this subsection.

\subsubsection{Step 1} 

\begin{itemize}
    \item (1) Each seller $i$ selects their secret trading data $S_i$; 
    \item (2) Each seller $i$  independently generates an encryption key $K_i$ for their secret data address; 
    \item (3) Each seller $i$ uploads their data to IPFS and obtains the corresponding IPFS content hash $ H{ipfs}_i$ provided by IPFS; 
    \item (4) Each seller  $i$ computes the encrypted address information $C_i=Encrypt(K_i, address_i)$; 
    \item (5) Each seller computes the hash of their encryption key $H(K_i)$; 
    \item (6) Each seller uses zk-SNARK to generate a proof $\pi_i$ demonstrating that they possess both $S_i$ and $K_i$, and each $H(K_i)$ matches the given hash value, and $S_i$ satisfies $F()$; and each $C_i$ is the encryption of the address using $K_i$, and the associated data matches the contents referenced by the IPFS address.
\end{itemize}

\subsubsection{Step 2} 

\begin{itemize}
    \item (1) Each seller presents to the buyer their corresponding $H(K_i), C_i$,  IPFS content hash $H{ipfs}_i$ and proof $\pi_i$;

    \item (2) The buyer verifies the validity of each $\pi_i$ to ensure that the seller indeed possesses the dataset $S_i$ and the corresponding encryption key $K_i$; and $S_i$ satisfies $F()$; and $C_i$ is the encryption of the address using $K_i$; and the data is indeed stored at the corresponding IPFS address.
\end{itemize}

\subsubsection{Step 3}

\begin{itemize}
    \item (1) The buyer deploys a smart contract on the blockchain, depositing the payment amount $A_i$ for each $S_i$ and specifies the payment condition: each seller must provide a decryption key $K_i$ such that $C_i$ can be decrypted to reveal the data address, and the result matches the corresponding hash preimage;
    \item (2) The contract includes the following logic: i.  If the provided $K_i$ successfully decrypts $C_i$ and the resulting address satisfies the condition that $H(K_i)$ matches the expected value, the payment $A_i$ is released to the seller $i$; ii. Otherwise, after a predefined timeout, the payment $A_i$ is refunded to the buyer.
\end{itemize}

\subsubsection{Step 4}

\begin{itemize}
\item The sellers submit the set of decryption keys $[K_1, K_2, \cdots, K_i, \cdots K_n]$  to the smart contract;

\item The blockchain network verifies whether each $K_i$ can correctly decrypt the corresponding $C_i$ obtain the address, and whether $H(K_i)$ matches the expected hash. If all verifications pass, the smart contract releases the payment $A_i$ to the respective seller $i$.
\end{itemize}

\subsubsection{Step 5}

\begin{itemize}
    \item If the key set passes verification, the seller $i$ receives the payment amount $A_i$ and the buyer decrypt the $address_i$ using $Decrypt(K_i, C_i)$;

    \item If the verification fails or the seller fails to submit the key $K_i$ within the specified time, the payment $A_i$ is refunded to the buyer.
\end{itemize}

In general, only one of the key or the data is exposed, and the data address is privately shared. Privacy refers to the data.

\subsection{Properties of Our Yotta Scheme}

In the previous discussion, we presented the first scheme for large-scale one-to-many batch data trading. In practice, however, there may be multiple buyers, each independently publishing their own evaluation function $F$, and sellers are free to sell their data to multiple buyers. Therefore, our Yotta scheme essentially supports many-to-many large-scale data trading on blockchain systems. In the following, we revisit our proposed scheme to assess its compliance with the properties outlined in the SQUATS principles for Web 3.0.

\begin{itemize}
    \item \textbf{Scalability:} In traditional data trading models, transactions between buyers and sellers are typically limited to atomic exchanges. In contrast, our approach enables many-to-many large-scale data trading and is even scalable to future yottabyte-scale scenarios, as our protocol could enable users to sell their data in batches. Furthermore, by leveraging recursive SNARKs~\cite{boneh2020halo}, our protocol supports the aggregation of proofs from multiple sellers, allowing the buyer to verify all proofs in a single step, thereby significantly improving trading efficiency.
 
    \item \textbf{Quality:} In IPFS, each file is assigned a unique Content Identifier (CID), which is a cryptographic hash computed from the file's content. This content-based addressing ensures that identical files produce the same CID. In our protocol, we leverage zk-SNARKs to generate proofs that bind the CID to the actual data, demonstrating that the data stored in IPFS is exactly what passes the evaluation function. This not only guarantees data integrity but also enables easy verification of consistency and correctness during the data trading process. Thus, by leveraging zk-SNARKs, the proof $\pi$ can be used to verify whether the data satisfies the specified evaluation function $F()$, thereby enabling verification of data quality.

    \item \textbf{Usability:} Our scheme also supports convenient horizontal enhancements and extensions. It can provide user-friendly APIs to facilitate ease of use.

    \item \textbf{Autonomy:} Our approach integrates seamlessly with blockchain smart contracts, requiring minimal manual labor or intervention.

    \item \textbf{Transparency:} The protocol records all transactions on the blockchain, making the entire trading process fully transparent and auditable~\cite{chen2022blockchain}. All data hashes and zero-knowledge proofs are stored on-chain, ensuring that no party can tamper with the transaction history, thereby improving the trustworthiness and security of data trading for our Yotta.

    \item \textbf{Security:} In our scheme, we do not have an untrusted third party to access all the data.  In our design, only one of the keys or the data is publicly revealed, while the data address is privately shared. Privacy here refers to the data itself; only the encryption key $K_i$ is visible on the blockchain, preventing any third party from accessing the actual data content.

\end{itemize}

Based on the above discussion, we conclude that Yotta effectively satisfies all the essential properties required for large-scale data trading in the Web 3.0 era.

\subsection{Discussions of Yotta}

\subsubsection{Complexity} We then discuss the complexity of our Yotta scheme.

\textbf{Time Complexity:} Our scheme improves trading efficiency in the following ways.

\begin{itemize}

    \item Our approach ensures transparency and tamper-resistance by combining complete IPFS content addressing with zk-SNARK-based proofs, which feature constant proof size and computation cost that does not grow with the data volume.

    \item Proofs on the seller side: All required zero-knowledge proofs can be generated by the seller, including (1) Proof that the data passes the buyer's evaluation function. (2) Proof that the seller truly possesses the secret information in the dataset and that this information matches the content stored at the corresponding IPFS address. (3) Proof that the provided decryption key is indeed the correct key for decrypting the data address. This design shifts the computational burden away from the buyer, significantly reducing their time cost.
    
    \item Recursive SNARK Verification on the buyer side: We leverage recursive SNARKs to batch-verify proofs from multiple sellers, substantially lowering the computational resource consumption. This approach not only accelerates verification but also reduces its computational cost.

\end{itemize}

\textbf{Space Complexity:} Our scheme highlights the following methods to reduce communication overhead.

\begin{itemize}

    \item Proof size: The transmitted proof is succinct, requiring only a small and constant amount of data regardless of the input size. 
    
    \item Encrypted Data Storage: Sellers encrypt their data and store them on IPFS. Only the IPFS address and the corresponding decryption key need to be transmitted on blockchain systems, which significantly reduces the overhead of transferring large datasets.
\end{itemize}

Thus, the Yotta scheme holds significant promise for practical deployment, meeting all previously defined principles while maintaining both time and space efficiency.

\subsubsection{Extensibility of Yotta}
\textbf{Application-level}: Sellers can freely sell their secret information to multiple buyers in batches or sell multiple pieces of secret data to a single buyer as long as the seller properly defines the granularity of each secret item.
\textbf{Technical-level}: Our scheme can be easily integrated with other emerging techniques, such as evaluation functions based on Shapley value for more accurate data quality assessment, more efficient decentralized storage systems beyond IPFS, or improved ZKP mechanisms.

\section{Simulation Experiment}

\begin{figure}[!th]
\centering
\includegraphics[width=0.5\linewidth]{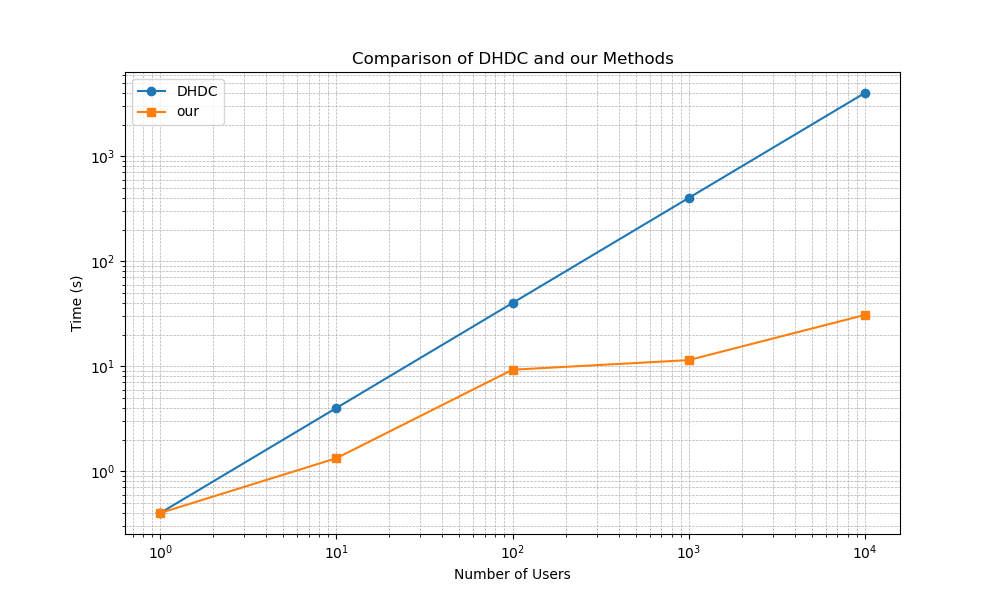}
\caption{Our simulation experiments with Yotta scheme prototype and DHDC.}
\label{fig:exp}
\end{figure}

As shown in Figure~\ref{fig:exp}, we implemented a prototype of our method, Yotta, and compared it with the most practical existing data-secure approach—DCDH. In our experiments, we have one buyer and multiple sellers. We gradually increased the number of participating sellers (users), with the y-axis plotted on a logarithmic scale.

It can be observed that with 10 users, our method achieves approximately 3× the performance of DCDH. With 100 users, the improvement reaches around 12×, and with 1,000 users, about 35×. Notably, at 10,000 users, our method outperforms DCDH by approximately 130×. These results strongly demonstrate the effectiveness of our approach compared to existing solutions. Our method demonstrates increasing performance gains as the number of users scales up, which aligns with realistic scenarios where thousands or even tens of thousands of sellers may participate.

Therefore, we present a practical, sublinear solution for large-scale batch data trading. Our method shows significant promise for future real-world applications.

\section{Application Discussions}

Our Yotta scheme can be deployed in various Web 3.0 applications:

In supply chain management, different participants (e.g., suppliers, manufacturers, retailers) can encrypt their data and store it in IPFS, enabling data exchange via blockchain. Each participant can verify the integrity and correctness of the data, ensuring trust and transparency across the supply chain. Specifically, (1) manufacturers can provide production data, which retailers and consumers can verify through the blockchain to ensure product quality and authenticity~\cite{li2022blockchain}; (2) logistics providers encrypt and upload transport data to IPFS. All supply chain nodes can verify this data to ensure a transparent and reliable delivery process. Our Yotta could easily meet the requirements of supply chain management.

In the data market, participants can publish data requirements along with corresponding evaluation methods. Individuals or organizations can encrypt compliant datasets and upload them to IPFS, enabling secure data sharing and trading via blockchain. Buyers can verify the integrity and correctness of the data before receiving the decryption key, ensuring data trustworthiness. In particular, (1) financial institutions can sell encrypted market data, customer behavior data, etc. Buyers verify the data’s integrity through proofs before obtaining the decryption keys; (2) hospitals and health research institutes can sell encrypted patient data. Researchers can purchase and verify the data before decrypting and using it~\cite{an2021secure}. Thus, our Yotta scheme is the ideal method that could fulfill all the demands of the data market.

\section{Conclusion}
Large-scale data trading is increasingly becoming a practical demand for participants in the Web 3.0 ecosystem. However, there is currently no practical solution available. In this work, we are the first to formalize the essential properties that a practical data trading scheme should satisfy, which we define as the SQUATS principles. Based on these principles, we design Yotta, a many-to-many large-scale data trading scheme. We provide a detailed explanation of Yotta’s design, analyze its features and extensibility, and implement a prototype system. Our evaluation shows that Yotta outperforms existing methods by up to 130×, demonstrating its effectiveness and practical potential.



\begin{thebibliography}{28}


\ifx \showCODEN    \undefined \def \showCODEN     #1{\unskip}     \fi
\ifx \showISBNx    \undefined \def \showISBNx     #1{\unskip}     \fi
\ifx \showISBNxiii \undefined \def \showISBNxiii  #1{\unskip}     \fi
\ifx \showISSN     \undefined \def \showISSN      #1{\unskip}     \fi
\ifx \showLCCN     \undefined \def \showLCCN      #1{\unskip}     \fi
\ifx \shownote     \undefined \def \shownote      #1{#1}          \fi
\ifx \showarticletitle \undefined \def \showarticletitle #1{#1}   \fi
\ifx \showURL      \undefined \def \showURL       {\relax}        \fi
\providecommand\bibfield[2]{#2}
\providecommand\bibinfo[2]{#2}
\providecommand\natexlab[1]{#1}
\providecommand\showeprint[2][]{arXiv:#2}

\bibitem[An et~al\mbox{.}(2021)]%
        {an2021secure}
\bibfield{author}{\bibinfo{person}{Baoyi An}, \bibinfo{person}{Mingjun Xiao}, \bibinfo{person}{An Liu}, \bibinfo{person}{Yun Xu}, \bibinfo{person}{Xiangliang Zhang}, {and} \bibinfo{person}{Qing Li}.} \bibinfo{year}{2021}\natexlab{}.
\newblock \showarticletitle{Secure crowdsensed data trading based on blockchain}.
\newblock \bibinfo{journal}{\emph{IEEE Transactions on Mobile Computing}} \bibinfo{volume}{22}, \bibinfo{number}{3} (\bibinfo{year}{2021}), \bibinfo{pages}{1763--1778}.
\newblock


\bibitem[Benet(2014)]%
        {benet2014ipfs}
\bibfield{author}{\bibinfo{person}{Juan Benet}.} \bibinfo{year}{2014}\natexlab{}.
\newblock \showarticletitle{Ipfs-content addressed, versioned, p2p file system}.
\newblock \bibinfo{journal}{\emph{arXiv preprint arXiv:1407.3561}} (\bibinfo{year}{2014}).
\newblock


\bibitem[Boneh et~al\mbox{.}(2020)]%
        {boneh2020halo}
\bibfield{author}{\bibinfo{person}{Dan Boneh}, \bibinfo{person}{Justin Drake}, \bibinfo{person}{Ben Fisch}, {and} \bibinfo{person}{Ariel Gabizon}.} \bibinfo{year}{2020}\natexlab{}.
\newblock \showarticletitle{Halo infinite: Recursive zk-SNARKs from any additive polynomial commitment scheme}.
\newblock \bibinfo{journal}{\emph{Cryptology ePrint Archive}} (\bibinfo{year}{2020}).
\newblock


\bibitem[Bowe et~al\mbox{.}(2020)]%
        {bowe2020zexe}
\bibfield{author}{\bibinfo{person}{Sean Bowe}, \bibinfo{person}{Alessandro Chiesa}, \bibinfo{person}{Matthew Green}, \bibinfo{person}{Ian Miers}, \bibinfo{person}{Pratyush Mishra}, {and} \bibinfo{person}{Howard Wu}.} \bibinfo{year}{2020}\natexlab{}.
\newblock \showarticletitle{Zexe: Enabling decentralized private computation}. In \bibinfo{booktitle}{\emph{2020 IEEE Symposium on Security and Privacy (SP)}}. IEEE, \bibinfo{pages}{947--964}.
\newblock


\bibitem[Chen et~al\mbox{.}(2023)]%
        {chen2023hyperplonk}
\bibfield{author}{\bibinfo{person}{Binyi Chen}, \bibinfo{person}{Benedikt B{\"u}nz}, \bibinfo{person}{Dan Boneh}, {and} \bibinfo{person}{Zhenfei Zhang}.} \bibinfo{year}{2023}\natexlab{}.
\newblock \showarticletitle{Hyperplonk: Plonk with linear-time prover and high-degree custom gates}. In \bibinfo{booktitle}{\emph{Annual International Conference on the Theory and Applications of Cryptographic Techniques}}. Springer, \bibinfo{pages}{499--530}.
\newblock


\bibitem[Chen et~al\mbox{.}(2022)]%
        {chen2022blockchain}
\bibfield{author}{\bibinfo{person}{Fei Chen}, \bibinfo{person}{Jiahao Wang}, \bibinfo{person}{Changkun Jiang}, \bibinfo{person}{Tao Xiang}, {and} \bibinfo{person}{Yuanyuan Yang}.} \bibinfo{year}{2022}\natexlab{}.
\newblock \showarticletitle{Blockchain based non-repudiable iot data trading: Simpler, faster, and cheaper}. In \bibinfo{booktitle}{\emph{IEEE INFOCOM 2022-IEEE Conference on Computer Communications}}. IEEE, \bibinfo{pages}{1958--1967}.
\newblock


\bibitem[Dai et~al\mbox{.}(2019)]%
        {dai2019sdte}
\bibfield{author}{\bibinfo{person}{Weiqi Dai}, \bibinfo{person}{Chunkai Dai}, \bibinfo{person}{Kim-Kwang~Raymond Choo}, \bibinfo{person}{Changze Cui}, \bibinfo{person}{Deiqing Zou}, {and} \bibinfo{person}{Hai Jin}.} \bibinfo{year}{2019}\natexlab{}.
\newblock \showarticletitle{SDTE: A secure blockchain-based data trading ecosystem}.
\newblock \bibinfo{journal}{\emph{IEEE Transactions on Information Forensics and Security}}  \bibinfo{volume}{15} (\bibinfo{year}{2019}), \bibinfo{pages}{725--737}.
\newblock


\bibitem[Dong et~al\mbox{.}(2018)]%
        {dong2018efficient}
\bibfield{author}{\bibinfo{person}{Xiangqian Dong}, \bibinfo{person}{Bing Guo}, \bibinfo{person}{Yan Shen}, \bibinfo{person}{Xuliang Duan}, \bibinfo{person}{YC Shen}, {and} \bibinfo{person}{H Zhang}.} \bibinfo{year}{2018}\natexlab{}.
\newblock \showarticletitle{An efficient and secure decentralizing data sharing model}.
\newblock \bibinfo{journal}{\emph{Chinese Journal of Computers}} \bibinfo{volume}{41}, \bibinfo{number}{5} (\bibinfo{year}{2018}), \bibinfo{pages}{1021--1036}.
\newblock


\bibitem[Entin(2023)]%
        {entin2023web3}
\bibfield{author}{\bibinfo{person}{Gregory Entin}.} \bibinfo{year}{2023}\natexlab{}.
\newblock \bibinfo{title}{Web3 Basics: The Future of a Decentralized Web}.
\newblock
\urldef\tempurl%
\url{https://www.linkedin.com/pulse/web3-basics-future-decentralized-web-gregory-entin-vsvuc/}
\showURL{%
\tempurl}
\newblock
\shownote{Accessed: 2025-04-28}.


\bibitem[Gabizon et~al\mbox{.}(2019)]%
        {gabizon2019plonk}
\bibfield{author}{\bibinfo{person}{Ariel Gabizon}, \bibinfo{person}{Zachary~J Williamson}, {and} \bibinfo{person}{Oana Ciobotaru}.} \bibinfo{year}{2019}\natexlab{}.
\newblock \showarticletitle{Plonk: Permutations over lagrange-bases for oecumenical noninteractive arguments of knowledge}.
\newblock \bibinfo{journal}{\emph{Cryptology ePrint Archive}} (\bibinfo{year}{2019}).
\newblock


\bibitem[IDC(2018)]%
        {idc2018datasphere}
\bibfield{author}{\bibinfo{person}{IDC}.} \bibinfo{year}{2018}\natexlab{}.
\newblock \bibinfo{title}{The Digitization of the World: From Edge to Core}.
\newblock
\urldef\tempurl%
\url{https://www.seagate.com/files/www-content/our-story/trends/files/idc-seagate-dataage-whitepaper.pdf}
\showURL{%
\tempurl}
\newblock
\shownote{Accessed: 2025-04-28}.


\bibitem[Jadbabaie(2014)]%
        {jadbabaie2014ieee}
\bibfield{author}{\bibinfo{person}{Ali Jadbabaie}.} \bibinfo{year}{2014}\natexlab{}.
\newblock \showarticletitle{IEEE Transactions on Network Science and Engineering}.
\newblock \bibinfo{journal}{\emph{IEEE Transactions on Network Science and Engineering}} \bibinfo{volume}{1}, \bibinfo{number}{01} (\bibinfo{year}{2014}), \bibinfo{pages}{2--9}.
\newblock


\bibitem[Jung et~al\mbox{.}(2017)]%
        {jung2017accounttrade}
\bibfield{author}{\bibinfo{person}{Taeho Jung}, \bibinfo{person}{Xiang-Yang Li}, \bibinfo{person}{Wenchao Huang}, \bibinfo{person}{Jianwei Qian}, \bibinfo{person}{Linlin Chen}, \bibinfo{person}{Junze Han}, \bibinfo{person}{Jiahui Hou}, {and} \bibinfo{person}{Cheng Su}.} \bibinfo{year}{2017}\natexlab{}.
\newblock \showarticletitle{Accounttrade: Accountable protocols for big data trading against dishonest consumers}. In \bibinfo{booktitle}{\emph{IEEE INFOCOM 2017-IEEE Conference on Computer Communications}}. IEEE, \bibinfo{pages}{1--9}.
\newblock


\bibitem[Jung et~al\mbox{.}(2018)]%
        {jung2018accounttrade}
\bibfield{author}{\bibinfo{person}{Taeho Jung}, \bibinfo{person}{Xiang-Yang Li}, \bibinfo{person}{Wenchao Huang}, \bibinfo{person}{Zhongying Qiao}, \bibinfo{person}{Jianwei Qian}, \bibinfo{person}{Linlin Chen}, \bibinfo{person}{Junze Han}, {and} \bibinfo{person}{Jiahui Hou}.} \bibinfo{year}{2018}\natexlab{}.
\newblock \showarticletitle{Accounttrade: Accountability against dishonest big data buyers and sellers}.
\newblock \bibinfo{journal}{\emph{IEEE Transactions on Information Forensics and Security}} \bibinfo{volume}{14}, \bibinfo{number}{1} (\bibinfo{year}{2018}), \bibinfo{pages}{223--234}.
\newblock


\bibitem[Krause(2024)]%
        {krause2024web3}
\bibfield{author}{\bibinfo{person}{David Krause}.} \bibinfo{year}{2024}\natexlab{}.
\newblock \showarticletitle{Web3 and the Decentralized Future: Exploring Data Ownership, Privacy, and Blockchain Infrastructure}.
\newblock \bibinfo{journal}{\emph{Privacy, and Blockchain Infrastructure (December 19, 2024)}} (\bibinfo{year}{2024}).
\newblock


\bibitem[Li et~al\mbox{.}(2022)]%
        {li2022blockchain}
\bibfield{author}{\bibinfo{person}{Chunlin Li}, \bibinfo{person}{SongYu Liang}, \bibinfo{person}{Jing Zhang}, \bibinfo{person}{Qiao-e Wang}, {and} \bibinfo{person}{Youlong Luo}.} \bibinfo{year}{2022}\natexlab{}.
\newblock \showarticletitle{Blockchain-based data trading in edge-cloud computing environment}.
\newblock \bibinfo{journal}{\emph{Information Processing \& Management}} \bibinfo{volume}{59}, \bibinfo{number}{1} (\bibinfo{year}{2022}), \bibinfo{pages}{102786}.
\newblock


\bibitem[Liu et~al\mbox{.}(2024)]%
        {liu2024pianist}
\bibfield{author}{\bibinfo{person}{Tianyi Liu}, \bibinfo{person}{Tiancheng Xie}, \bibinfo{person}{Jiaheng Zhang}, \bibinfo{person}{Dawn Song}, {and} \bibinfo{person}{Yupeng Zhang}.} \bibinfo{year}{2024}\natexlab{}.
\newblock \showarticletitle{Pianist: Scalable zkrollups via fully distributed zero-knowledge proofs}. In \bibinfo{booktitle}{\emph{2024 IEEE Symposium on Security and Privacy (SP)}}. IEEE, \bibinfo{pages}{1777--1793}.
\newblock


\bibitem[Liu et~al\mbox{.}(2021)]%
        {liu2021zkcnn}
\bibfield{author}{\bibinfo{person}{Tianyi Liu}, \bibinfo{person}{Xiang Xie}, {and} \bibinfo{person}{Yupeng Zhang}.} \bibinfo{year}{2021}\natexlab{}.
\newblock \showarticletitle{zkCNN: Zero knowledge proofs for convolutional neural network predictions and accuracy}. In \bibinfo{booktitle}{\emph{Proceedings of the 2021 ACM SIGSAC Conference on Computer and Communications Security}}. \bibinfo{pages}{2968--2985}.
\newblock


\bibitem[{Mozilla Foundation}(2024)]%
        {mozilla2024internet}
\bibfield{author}{\bibinfo{person}{{Mozilla Foundation}}.} \bibinfo{year}{2024}\natexlab{}.
\newblock \bibinfo{title}{Internet Health Report}.
\newblock
\urldef\tempurl%
\url{https://foundation.mozilla.org/en/insights/internet-health-report}
\showURL{%
\tempurl}
\newblock
\shownote{Accessed: 2025-04-28}.


\bibitem[Nakamoto(2008)]%
        {nakamoto2008bitcoin}
\bibfield{author}{\bibinfo{person}{Satoshi Nakamoto}.} \bibinfo{year}{2008}\natexlab{}.
\newblock \showarticletitle{Bitcoin: A peer-to-peer electronic cash system}.
\newblock  (\bibinfo{year}{2008}).
\newblock


\bibitem[Rosenberg et~al\mbox{.}(2023)]%
        {rosenberg2023zk}
\bibfield{author}{\bibinfo{person}{Michael Rosenberg}, \bibinfo{person}{Jacob White}, \bibinfo{person}{Christina Garman}, {and} \bibinfo{person}{Ian Miers}.} \bibinfo{year}{2023}\natexlab{}.
\newblock \showarticletitle{zk-creds: Flexible anonymous credentials from zksnarks and existing identity infrastructure}. In \bibinfo{booktitle}{\emph{2023 IEEE Symposium on Security and Privacy (SP)}}. IEEE, \bibinfo{pages}{790--808}.
\newblock


\bibitem[Speaker(2025)]%
        {opencamp2025web3talk}
\bibfield{author}{\bibinfo{person}{Unknown Speaker}.} \bibinfo{year}{2025}\natexlab{}.
\newblock \bibinfo{title}{Web3 \& Decentralization}.
\newblock \bibinfo{howpublished}{Talk at Bratislava OpenCamp 2025}.
\newblock
\urldef\tempurl%
\url{https://pretalx.opencamp.sk/bratislava-opencamp-2025/talk/A8EJ8V/}
\showURL{%
\tempurl}
\newblock
\shownote{Accessed: 2025-04-28}.


\bibitem[Taleb et~al\mbox{.}(2018)]%
        {taleb2018big}
\bibfield{author}{\bibinfo{person}{Ikbal Taleb}, \bibinfo{person}{Mohamed~Adel Serhani}, {and} \bibinfo{person}{Rachida Dssouli}.} \bibinfo{year}{2018}\natexlab{}.
\newblock \showarticletitle{Big data quality assessment model for unstructured data}. In \bibinfo{booktitle}{\emph{2018 International Conference on Innovations in Information Technology (IIT)}}. IEEE, \bibinfo{pages}{69--74}.
\newblock


\bibitem[Wood et~al\mbox{.}(2014)]%
        {wood2014ethereum}
\bibfield{author}{\bibinfo{person}{Gavin Wood} {et~al\mbox{.}}} \bibinfo{year}{2014}\natexlab{}.
\newblock \showarticletitle{Ethereum: A secure decentralised generalised transaction ledger}.
\newblock \bibinfo{journal}{\emph{Ethereum project yellow paper}} \bibinfo{volume}{151}, \bibinfo{number}{2014} (\bibinfo{year}{2014}), \bibinfo{pages}{1--32}.
\newblock


\bibitem[Yue et~al\mbox{.}(2017)]%
        {yue2017big}
\bibfield{author}{\bibinfo{person}{Li Yue}, \bibinfo{person}{Huang Junqin}, \bibinfo{person}{Qin Shengzhi}, {and} \bibinfo{person}{Wang Ruijin}.} \bibinfo{year}{2017}\natexlab{}.
\newblock \showarticletitle{Big data model of security sharing based on blockchain}. In \bibinfo{booktitle}{\emph{2017 3rd International Conference on Big Data Computing and Communications (BIGCOM)}}. IEEE, \bibinfo{pages}{117--121}.
\newblock


\bibitem[Zheng et~al\mbox{.}(2020)]%
        {zheng2020blockchain}
\bibfield{author}{\bibinfo{person}{Shuli Zheng}, \bibinfo{person}{Lixuan Pan}, \bibinfo{person}{Donghui Hu}, \bibinfo{person}{Meng Li}, {and} \bibinfo{person}{Yuqi Fan}.} \bibinfo{year}{2020}\natexlab{}.
\newblock \showarticletitle{A blockchain-based trading platform for big data}. In \bibinfo{booktitle}{\emph{IEEE INFOCOM 2020-IEEE Conference on Computer Communications Workshops (INFOCOM WKSHPS)}}. IEEE, \bibinfo{pages}{991--996}.
\newblock


\bibitem[Zheng et~al\mbox{.}(2017)]%
        {zheng2017overview}
\bibfield{author}{\bibinfo{person}{Zibin Zheng}, \bibinfo{person}{Shaoan Xie}, \bibinfo{person}{Hongning Dai}, \bibinfo{person}{Xiangping Chen}, {and} \bibinfo{person}{Huaimin Wang}.} \bibinfo{year}{2017}\natexlab{}.
\newblock \showarticletitle{An overview of blockchain technology: Architecture, consensus, and future trends}. In \bibinfo{booktitle}{\emph{2017 IEEE international congress on big data (BigData congress)}}. Ieee, \bibinfo{pages}{557--564}.
\newblock


\bibitem[Zyskind et~al\mbox{.}(2015)]%
        {zyskind2015decentralizing}
\bibfield{author}{\bibinfo{person}{Guy Zyskind}, \bibinfo{person}{Oz Nathan}, {et~al\mbox{.}}} \bibinfo{year}{2015}\natexlab{}.
\newblock \showarticletitle{Decentralizing privacy: Using blockchain to protect personal data}. In \bibinfo{booktitle}{\emph{2015 IEEE security and privacy workshops}}. IEEE, \bibinfo{pages}{180--184}.
\newblock


\end{thebibliography}



\end{document}